# Skill Transfer System that Visualizes and Presents Tactile Information in an AR Environment

Takaya Nida[1], Masamune Waga[1], Yuta Hamada[1], Masashi Konyo[1], Yushi Nakaya[2],
Shubhamkumar Pandey[2], and Satoshi Tadokoro[1]

[1] *Graduate School of Information Sciences. Tohoku University, Japan*

[2] *Adansons Inc., 1-4-9, Aobaku Kokubuncho, Sendai City, Miyagi, Japan (980-0803)*

(Email: nida.takaya@rm.is.tohoku.ac.jp)

**Abstract ---** In recent years, the lack of successors for traditional skills has become an issue. To solve this problem, we propose a skill transfer system that presents tactile information in spatial tasks as a color map on an AR space. We believe that providing the operator with feedback of the force and tactile information during the work is useful for learning skills that require time to master. Furthermore, by following the operator's hand and presenting tactile information, we expect to accelerate the learning of skills by not only presenting tactile information as a physical sensation, but also by making the operator associate tactile information with position.

## 1 INTRODUCTION

In recent years, the worldwide aging of society has led to a serious shortage of successors to traditional skills and other technical skills. It is believed that it takes many years to acquire these skills, and there is a need to speed up the process to preserve excellent skills for future generations. To solve this problem, remote skills education using visualization technologies such as VR, AR, and projection mapping is attracting attention.

Studies combining these visualization techniques with tactile stimuli have been reported. Examples include HaptoMapping, a visual and haptic augmented reality system that can consistently present independent visual and haptic content on a physical surface, and SoftAR, an AR technology based on a pseudo-haptic mechanism that visually manipulates the sensation a user has when pushing a soft object [1][2]. These systems map tactile information to visual information by adding visual information to tactile information, but they cannot be used as-is for tasks such as cutting that require skilled artistry. In VR, the user learns skills in a pre-modeled workspace, which may cause a sensory discrepancy with the real world.

The force and tactile information generated during actual work is recorded by the sensor and fed back to the operator, which is considered useful for learning skills. To measure force, it is necessary to embed a sensor in the work object or attach it to the end of a tool. However, there are many restrictions because sensors cannot be attached to some work objects, and even if they are connected to tools, they may interfere with the actual work. Vibration sensors can measure vibrations propagated near the work area or on the human body, making them applicable to a wide range of tasks.

In this study, we propose a skill transfer system that presents tactile information in spatial tasks as a color map on an AR space. The operator wears a bracelet-type tactile device with a built-in sensor and vibrator, and the tactile information propagated to the operator is recorded. The tactile information is presented as a color map following the operator's hand by motion capture, and the operator can learn the skill by associating the tactile information with his/her position. In addition, by generating images superimposed on the real workspace as AR, it is thought that there will be no sensory gap with the real world. Furthermore, tactile information is fed back not only as images but also as vibrations on the wrist.

The purpose of this research is "to construct a skill transfer system that visualizes and presents tactile information at the operator's hand in an AR space in order to promote the operator's skill learning.

## 2 Method

### 2.1 Concept of AR tactile visualization system

In spatial tasks, the AR haptic visualization system uses 3D haptic visualization and vibration feedback to provide a clearer representation of the movement experience. We believe that this makes it easier to learn skills that are difficult to grasp using haptic feedback alone. In this study, we propose a system that allows beginners to acquire skills by comparing their own sensations with those of skilled users and visualizing their sensations in real time using a color map.

Generally, when visualizing real-time information, data is displayed in the form of graphs. However, in this case, the user needs to check the graph while performing the task, making it difficult to concentrate fully on the task. In order to provide simple visualization and maintain concentration, this system displays information using a color map.

The experience can be recorded in advance and played back later. The data includes vibration feedback during the task and a color-mapped video from the operator's perspective. The operator can play this data offline to check their own movements and compare their own sensations with the movements of the expert, allowing them to learn through repetition.

### 2.2 Intensity Segment Modulation (ISM)

The authors have proposed Intensity Segment Modulation (ISM) as a technique for converting a vibration signal into an amplitude-modulated wave of an arbitrary carrier frequency while maintaining the sensation of vibration. ISM is a conversion technique that focuses on two perceptual characteristics of human high-frequency vibration: intensity perception and envelope perception[3]. High-frequency vibration is divided into segments of 5 ms, and converted into amplitude modulation waves of their own frequency while retaining the calculated perceptual intensity. The waveform is decomposed into frequency-specific waveforms using Empirical Mode Decomposition (hereafter EMD), and the total of the calculated perceptual intensities is taken. The definition formula used to calculate the perceptual intensity is shown in Equation (1).

$$I(f) = \left[\left(\frac{A_T(f)}{A(f)}\right)^2\right]^{\alpha(f)} \quad (1)$$

$A(f)$ is the amplitude of vibration, f is the frequency, $I(f)$ and $\alpha$ are the vibration discrimination threshold and the index of vibration intensity, which depend on the frequency f. The vibrotactile stimuli used in this study are amplitude-modulated waves generated by ISM. The frequency of the amplitude-modulated waves used in this paper is 200 Hz.

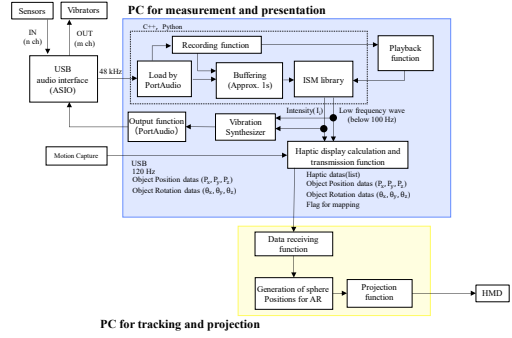

Figure 1. Block Diagram

Furthermore, the perceptual intensity calculated during the ISM conversion process was used as a parameter for the color map. By referring to values based on human perceptual characteristics and visualizing them, it is hoped that it will be possible to grasp the sensations of tasks that cannot be understood by just experiencing the sensation.

## 3 SYSTEM IMPLEMENTATION

### 3.1 Measurement of tactile information and color mapping

Figure 1 shows an overview of the implemented system. A bi-directional arm-worn haptic device is used as a device for measuring and presenting vibrations. This device has four units arranged in a ring at equal angles. Each unit is equipped with a vibration sensor (TOKIN, VS-BV201) and a vibrator (Nidec, Slider1), and by arranging the four units at equal intervals on the wrist, it is possible to reproduce the spatial distribution of vibrations. The vibrations measured on the wrist are connected to a PC for vibration measurement and presentation via USB audio (Roland, Octa-Capture UA1010) and recorded. After the recorded vibration data is buffered, the ISM conversion library is used to calculate the perceived intensity and low-frequency components below 100Hz from the waveform of each buffer. The perceived intensity used in the color map is the average of the perceived intensity measured at four points.

Motion capture (Optitruck, V120: Duo) reads the position information of the work tool. By attaching four reflective markers to the work tool, the rigid position and angle information of the tool is read by 3D measurement and reflected in the Unity virtual space. A spherical object is generated in Unity, and the position and posture are calibrated so that the spherical object is at the tip of the tool.

The vibration information captured on the PC for measurement is sent to the PC for visualization via socket communication. The received perceptual intensity (hereafter referred to as 'intensity') is sent to the PC for visualization, and colors are assigned to the spheres based on the intensity values. In this method, the maximum intensity value received in advance is set, and the intensity is normalized based on that value. If the received intensity exceeds the maximum value, the value is specified as 1. The color map is then realized by assigning the normalized values to rgb values. The color map used is Turbo, which was announced by Google. Turbo is a color map that improves visibility and takes into account colorblindness. By applying a process that duplicates the spheres with this color map, the trajectory of the movement is expressed. As an AR device, we use an optical see-through type device. The optical see-through type is a method that overlays AR objects directly on the display, so it is less likely to cause a sensory gap for the user. In this method, we use an HMD (HoloLens2, Microsoft) of the same see-through type.

### 3.2 Vibration presentation

Using the perceived intensity and low-frequency information calculated by the ISM conversion library, a vibration waveform with a carrier frequency of 200 Hz is synthesized with the same perceived intensity as the original waveform. After signal processing, the waveform is amplified via USB audio and amplifier (Syntacts, V3.1) and output to a bi-directional bracelet-type sensory transmission device. The operator can simultaneously experience his/her own motion and visualize tactile information in real-time. Furthermore, the vibration information and the work video can be measured, allowing the operator to confirm his or her own movements during work.

### 4 CONCLUSION

In this study, we proposed a skill transfer system that visualizes tactile information and presents it in an AR environment. In the future, we plan to conduct experiments that will allow us to quantitatively evaluate the effectiveness of the system. Specifically, in order to confirm whether the visualization of tactile information is effective for skill transfer, we will ask the participants to perform the same task under four conditions: 'visualization using color ', 'visualization using color maps only', 'tactile sensation only', and 'no information presented', and we will ask them to evaluate how easy it was to reproduce the movements in each condition using a questionnaire. In our demonstration, we will ask participants to actually wear the HMD and experience the AR visualization

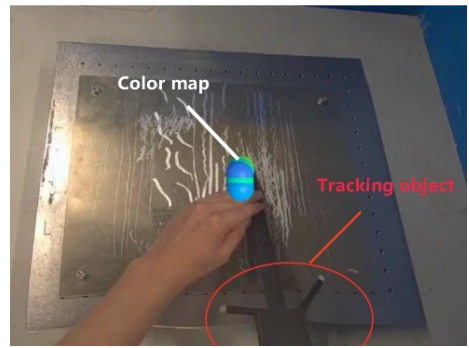

Figure 2. Color map generated by the developed the system

### ACKNOWLEDGEMENT

The results were obtained as a result of work commissioned by the New Energy and Industrial Technology Development Organization (NEDO) (JPNP21004).